\documentclass[preprint,eqsecnum,preprintnumbers,nofootinbib,byrevtex,prd,aps,showpacs,showkeys,groupedaddress,floatfix]{revtex4}
\usepackage{bm}
\usepackage{graphics}
\usepackage{graphicx}
\usepackage{epsfig}
\usepackage{amssymb}
\usepackage{amsmath}
\begin{document}

\title{Inclusive Gluon Production in Pion-Proton Collisions and the Principle Maximum Conformality Renormalization Scale}
\author{A.~I.~Ahmadov$^{1,2}$~\footnote{E-mail: ahmadovazar@yahoo.com}}
\author{C.~Aydin$^{1,3}$~\footnote{E-mail: coskun@ktu.edu.tr}}
\author{O.~Uzun$^{1}$~\footnote{E-mail: $\mbox{oguzhan}_-\mbox{deu@hotmail.com}$}}
\affiliation {$^{1}$ Department of Physics, Karadeniz Technical
University, 61080, Trabzon, Turkey \\
$^{2}$ Department of Theoretical Physics, Baku State University, Z.
Khalilov st. 23, AZ-1148, Baku, Azerbaijan\\
$^{3}$ Department of Physics, University of Surrey, Guildford Surrey
GU2 7XH, United Kingdom}

\date{\today}

\begin{abstract}
The contribution of the higher-twist mechanism to the large-$p_T$
inclusive gluon production cross section in $\pi p$ collisions is
calculated in case of the principle of maximum conformality and
Brodsky-Lepage-Mackenzie approaches in the dependence of the pion
distribution amplitudes. The higher-twist cross sections obtained in
the framework of the principle of maximum conformality and
Brodsky-Lepage-Mackenzie approaches, and compared and analyzed in
relation to the leading-twist cross sections and each other.
\end{abstract}

 \maketitle

\section{\bf Introduction}
It is well know that Quantum chromodynamics(QCD) is the fundamental
theory of the strong interactions. Many researchers study QCD to
describe the structure and dynamics of hadrons at the amplitude
level. The hadronic distribution amplitude in terms of internal
structure degrees of freedoms is important in QCD process
predictions.

One of the basic problems is choosing the renormalization scale in
running coupling constant $\alpha_{s}(Q^2)$. It is already stated in
Ref.~\cite{Brodsky1} that in perturbative QCD (pQCD) calculations,
the argument of the running coupling constant in both the
renormalization and factorization scale $Q^2$ should be taken equal
as to the square of the momentum transfer of a hard gluon in a
corresponding Feynman diagram. However, in this definition infrared
singularity is removed in $\alpha_{s}({Q}^2)$. Optimal approaches
for solution of this problem can be found with the
Brodsky-Lepage-Mackenzie (BLM)~\cite{Brodsky1} and the Principle of
Maximum Conformality (PMC)~\cite{Brodsky10} methods.

In the perturbative QCD, the physical information of the inclusive
gluon production is obtained efficiently; therefore, it can be
directly compared to the experimental data.

Using the frozen and running coupling constant approaches the higher
twist effects were already calculated by many authors
~\cite{Bagger,Bagger1,Baier,Sadykhov,Ahmadov1,Ahmadov2,
Ahmadov3,Ahmadov4,Ahmadov5,Ahmadov6,Ahmadov7,Ahmadov8,Ahmadov9}.

The calculation and  analysis of the higher-twist effects on the
dependence of the pion distribution amplitude in inclusive gluon
production at $\pi p$ collision within PMC and BLM approaches are
important research problem. In this work we computed the
contribution of the higher-twist effects to an inclusive gluon
production cross section by using various pion distribution
amplitudes from holographic and perturbative QCD. We have also
estimated and performed comparisons of the leading and the
higher-twist contributions.

The mechanism for choosing renormalization scale is provided in
Sec.~\ref{ir}. Some formulas for the higher-twist and leading-twist
cross sections are presented in Sec.~\ref{ht}, and the numerical
results for the cross section and discussion of the dependence of
the cross section on the pion distribution amplitudes are provided
in Sec.~\ref{results}. Finally, our conclusions and the highlights
of the study are listed in Sec.~\ref{conc}.

\section{CHOOSING THE RENORMALIZATION SCALE USING THE PRINCIPLE OF MAXIMUM CONFORMALITY}
\label{ir}

In principle, all measurable quantities in QCD should  be invariant
under any choice of renormalization scale and scheme. It is clear
that the use of different scales and schemes may lead to different
theoretical predictions.  Taking into account this fact the
constructive mathematical apparatus for defining QCD is a choice of
the  renormalization scale which makes scheme independent results at
all fixed order in running coupling constant $\alpha_{s}$.

The main idea of PMC/BLM,
proposed and being developed by Brodsky {\it et
al.}~\cite{Brodsky10,Brodsky9,Brodsky5,Brodsky6,Brodsky7,Brodsky8},
is that after proper procedures, all nonconformal ${\beta_{i}}$
terms in the perturbative expansion are collected into the running
coupling so that the remaining terms in the perturbative series are
identical to those of a conformal theory, namely the corresponding
theory with ${\beta_{i}} = {0}$. Then the QCD predictions from
PMC/BLM are independent of renormalization scheme. It has been found
that PMC/BLM satisfies all self-consistent conditions~\cite{Lu}. As
analyzed in Ref.~\cite{Brodsky9}, after PMC/BLM scale setting, the
divergent renormalon series $(n!\beta_{i}^{n}\alpha_{s}^{n})$ does
not appear in the conformal series.

Usually in  QCD calculations, one chooses the renormalization scale
$\mu^{2}$ equal to $ Q^2, p_{T}^2,....p_{T}^2/2$,  where $Q^2$ is a
typical momentum transfer in the process obtained directly from
Feynman diagram and $p_{T}^2$ is the squared transfer momentum of
the observed particle. This approach has a problem, predicting QCD
cross sections becoming negative at next-to-leading
order~\cite{Brodsky10,Maitre}.

The renormalization scale in QED, for example in the modified
minimal subtraction ($\overline{MS}$) scheme, has the
form~\cite{Brodsky10},
\begin{equation}
\mu_{\overline{MS}}^2=Q^2e^{-5/3}
\end{equation}
if $Q^2=-q^2$ is spacelike.

Therefore, we can choose
\begin{equation}
\alpha_{\overline{MS}}(q^2e^{-5/3})=\alpha_{GM-L}(q^2).
\end{equation}

Such an analogy  of  renormalization scales between the
$\overline{MS}$ and Gell-Mann-Low schemes leads to the displacement
of these schemes with $e^{-5/3}$ factor. It was chosen to determine
and provide the minimal dimensional regularization scheme
~\cite{Bardeen,Brodsky10}.

Thus the PMC scale for the calculation differential cross section in
the $\overline{MS}$ scheme is given simply by the $\overline{MS}$
scheme displacement of the gluon virtuality as
$\mu_{PMC}^{2}=e^{-5/3}Q^2$.

The PMC method is a general approach to set the renormalization
scale in QCD, including purely gluonic processes. It is scheme
independent and
avoids the
renormalon growth due to the absence of the
$\beta$ function terms in the perturbative expansion.
Since the
$\beta$-function in QCD is gauge invariant in any correct
renormalization scheme, the resulting conformal series is also gauge
invariant. Thus, the PMC is a gauge-invariant procedure.

For the calculation of QCD processes various schemes  are used, such
as minimal subtraction (MS), modified minimal subtraction
$\overline{(MS)}$, momentum subtraction (MOM) or the BLM scheme. In
the higher-twist processes, one of the main problem is choosing
renormalization scale. In this paper, for the fixing renormalization
scale, PMC method is used  and compared with the BLM method.

\section{HIGHER-TWIST AND LEADING-TWIST CONTRIBUTIONS TO INCLUSIVE GLUON PRODUCTION}
\label{ht}

The higher-twist Feynman diagrams for the inclusive gluon production
in the pion-proton collision $\pi p \to g X$ are shown in Fig.1. The
amplitude for this subprocess is found by means of the
Brodsky-Lepage formula~\cite{Lepage2}
\begin{equation}
M(\hat s,\hat
t)=\int_{0}^{1}{dx_1}\int_{0}^{1}dx_2\delta(1-x_1-x_2)\Phi_{M}(x_1,x_2,Q^2)T_{H}(\hat
s,\hat t;x_1,x_2)
\end{equation}
where $T_H$ is  the sum of the graphs contributing to the
hard-scattering part of the subprocess. For the  higher-twist, the
subprocess $\pi q_{p} \to g q$ is taken, which contributes to $\pi p
\to g X$,  where $q_{p}$ is a constituent of the initial proton
target. As seen from Fig.1, processes  $\pi^{+} p \to g X$ and
$\pi^{-} p \to g X$ arise from subprocesses as $\pi^{+} d_{p} \to g
u$ and $\pi^{-} u_{p} \to g d$, respectively.

We calculate the inclusive gluon production higher-twist cross
section for various pion distribution amplitudes. Therefore, the
higher-twist subprocess $\pi p \to g X$ is incorporated into the
full inclusive cross section.

The production of the hadronic gluon or jets in the large transverse
momentum  is available at the high energy, especially at the Large
Hadron Collider. Hadronic gluon as final are a product of the
hard-scattering processes, before hadronization. In the final state
this hadronic gluon or jets are fragmented or converted to hadron.
Dynamical properties  of the jet is close to the parent parton,
which carried one of the part four momentum of parent parton.
Therefore, to get closer  to the underlying parton level kinematics
the gluon production process should be studied~\cite{Owens}.

We can write the higher-twist cross section as

\begin{equation}
E\frac{d\sigma}{d^{3}p}(\pi p \to g X )=\int_{0}^{1}dx \delta(\hat
s+\hat t+\hat u)\hat s G_{q/{p}}(x, Q^2)\frac
{1}{\pi}\frac{d\sigma}{d\hat t}(\pi q_{p} \to g q),
\end{equation}
where $G_{q/p}(x, Q^2)$ is the quark distribution function  inside
a proton.

The Mandelstam invariant variables for higher-twist subprocess $\pi
q_{p} \to g q$ we can write in the form:
\begin{equation}
\hat s=(p_1+p_{g})^2=(p_2+p_{\pi})^2,\quad \hat
t=(p_{g}-p_{\pi})^2,\quad \hat u=(p_1-p_{\pi})^2.
\end{equation}
and from ~\cite{Berger}
\begin{equation}
\frac{d\sigma}{d\hat t}(\hat s,\hat t,\hat u)=\frac {256\pi^2}
{81{\hat s}^2}\left[D(\hat s,\hat u)\right]^2\left(-\frac{\hat
t}{{\hat s}^2}-\frac{\hat t}{{\hat u}^2}\right)\, ,
\end{equation}
where
\begin{equation}
D(\hat s,\hat u)=\int_{0}^{1}dx
\alpha_{s}^{3/2}(Q_1^2)\left[\frac{\Phi_{\pi}(x,Q_{1}^2)}{x(1-x)}\right]+\int_{0}^{1}dx
\alpha_{s}^{3/2}(Q_2^2)\left[\frac{\Phi_{\pi}(x,Q_{2}^2)}{x(1-x)}\right],
\end{equation}
$Q_{1}^2=(1-x)\hat s$ and $Q_{2}^2=-x\hat u$  represent the momentum
square of the hard gluon in Fig.1. So, in the BLM approach
renormalization scales are $\mu_{1}^{BLM}=Q_{1}$ and
$\mu_{2}^{BLM}=Q_{2}$, but in the PMC approach they are
$\mu_{1}^{PMC}=Q_{1}e^{-5/6}$ and $\mu_{2}^{PMC}=Q_{2}e^{-5/6}$.

There are many forms of pion distribution amplitude available in the
literature. In our numerical calculations, we used several choices,
such as the asymptotic distribution amplitude derived in pQCD
evalution~\cite{Lepage1}, the
Vega-Schmidt-Branz-Gutsche-Lyubovitskij (VSBGL) distribution
amplitude~\cite{ Vega},
 distribution amplitudes predicted by
 AdS/CFT~\cite{Brodsky2,Brodsky3}, and the Chernyak-Zhitnitsky(CZ)~\cite{chernyak}, the
Bakulev-Mikhailov-Stefanis (BMS)~\cite{Bakulev} and twist-three
distribution amplitudes (HW)~\cite{Huang}:
\begin{equation}
\Phi_{asy}(x)=\sqrt{3}f_{\pi}x(1-x),
\label{asy}
\end{equation}
\begin{equation}
\Phi_{VSBGL}^{hol}(x)=\frac{A_1k_1}{2\pi}\sqrt{x(1-x)}exp\left(-\frac{m^2}{2k_{1}^2x(1-x)}\right),
\end{equation}
\begin{equation}
\Phi^{hol}(x)=\frac{4}{\sqrt{3}\pi}f_{\pi}\sqrt{x(1-x)},
\end{equation}
\begin{equation}
\Phi_{CZ}(x,\mu_{0}^2)=\Phi_{asy}(x)\left[C_{0}^{3/2}(2x-1)+\frac{2}{3}C_{2}^{3/2}(2x-1)\right],
\end{equation}
\begin{equation}
\Phi_{BMS}(x,\mu_{0}^2)=\Phi_{asy}(x)\left[C_{0}^{3/2}(2x-1)+0.20C_{2}^{3/2}(2x-1)-0.14C_{4}^{3/2}(2x-1)\right],
\end{equation}
\begin{equation}
\Phi_{HW}(x)=\frac{A_p\beta^2}{2\pi^2}\left[1+B_pC_{2}^{1/2}(2x-1)+C_pC_{4}^{1/2}(2x-1)\right]exp\left(-\frac{m^2}{8\beta^2
x(x-1)}\right),
\label{HW}
\end{equation}
$C_{n}^{\lambda}(2x-1)$ are Gegenbauer polynomials  with the
recurrence relation
\begin{equation}
nC_{n}^{(\lambda)}(\xi)=2(n+\lambda-1)\xi
C_{n-1}^{(\lambda)}(\xi)-(n+2\lambda-2)C_{n-2}^{(\lambda)}(\xi).
\end{equation}
The pion distribution amplitude can be expanded as
\begin{equation}
\Phi_{\pi}(x,Q^2)=\Phi_{asy}(x)\left[1+\sum_{n=2,4..}^{\infty}a_{n}(Q^2)C_{n}^{3/2}(2x-1)\right].
\end{equation}
Finally, differential  cross section for the process $\pi p \to g X$
is written as ~\cite{Owens}
\begin{equation}
E\frac{d\sigma}{d^{3}p}(\pi p\to g X )=\frac{s}{s+u} xG_{q/p}(x,
Q^2)\frac {256\pi} {81{\hat s}^2}\left[D(\hat s,\hat
u)\right]^2\left(-\frac{\hat t}{{\hat s}^2}-\frac{\hat t}{{\hat
u}^2}\right).
\end{equation}
It should be noted that, as seen from Eq.(3.4) and Eq.(3.14), the
higher-twist cross section is linear with respect to $\hat t$, so
the cross section vanishes if the scattering angle between the final
gluon and incident pion is approximately equal to zero, $\theta=0$.
Also, as seen from Eq.(3.4), the higher-twist cross section
proportional to $\hat s^{-3}$, which shows that higher-twist
contributions to the $\pi p\to gX$ cross section have the form of
$p_{T}^{-6}f(x_{F},x_{T})$.

The difficulty comes in extracting the higher-twist corrections to
the inclusive gluon production cross section. One can also consider
the comparison of higher-twist corrections with leading-twist
contributions. For the leading-twist subprocess in the inclusive
gluon production, we take $q\bar{q} \to g\gamma$ as a subprocess of
the quark-antiquark annihilation. The differential  cross section
for subprocess $q\bar{q} \to g\gamma$ is
\begin{equation}
\frac{d\sigma}{d\hat t}(q\bar{q} \to g\gamma)=\frac{8}{9}\pi\alpha_E
\alpha_s(Q^2)\frac{e_{q}^2}{{\hat s}^2}\left(\frac{\hat t}{\hat
u}+\frac{\hat u}{\hat t}\right).
\end{equation}
As seen from Eq.(3.15), the leading-twist is proportional to
$p_{T}^{-4}$.

The leading-twist  cross section for production of inclusive gluon
is~\cite{Berger1}
\begin{equation}
\Sigma_{M}^{LT}\equiv E\frac{d\sigma}{d^{3}p}(\pi p \to g X
)=\int_{0}^{1}dx_{1} \int_{0}^{1}dx_{2} \delta(\hat s+\hat t+\hat u)
G_{\overline{q}/{M}}(x_{1},Q_{1}^2)G_{q/{p}}(x_{2},Q_{2}^2)\frac
{\hat s}{\pi}\frac{d\sigma}{d\hat t}(q\bar{q} \to g\gamma),
\end{equation}
where
$$
 \hat{s}=x_{1}x_{2}s,\ \hat{t}=x_{1}t,\  \hat{u}=x_{2}u.
$$
The leading-twist contribution to the large-$p_{T}$ gluon production
cross section in the process $\pi p\to g X$ is
\begin{equation}
\Sigma_{M}^{LT}\equiv E\frac{d\sigma}{d^{3}p}(\pi p \to g X
)=\int_{0}^{1}dx_{1} \frac{1}{x_1s+u}
G_{\overline{q}/{M}}(x_{1},Q_{1}^2)G_{q/{p}}(1-x_{1},Q_{2}^2)\frac
{\hat s}{\pi}\frac{d\sigma}{d\hat t}(q\bar{q} \to g\gamma).
\end{equation}

We denote  higher-twist cross section calculated with the PMC and
BLM approaches by $(\Sigma_{g}^{HT})_{PMC}$ and
$(\Sigma_{g}^{HT})_{BLM}$,
 respectively.

\section{NUMERICAL RESULTS AND DISCUSSION}
\label{results}

We will discuss for the higher-twist contribution in the process of
inclusive gluon production in $\pi p$ collisions. In the numerical
calculations  for the fixing renormaliation scale, we applied the
PMC and BLM approaches. For $\pi^{+} p \to g X$ and $\pi^{-} p \to g
X$ processes, we take $\pi^{+} d_{p} \to g u$, and $\pi^{-} u_{p}
\to g d$  as respective subprocesses.

For the dominant leading-twist subprocess for the gluon production,
the quark-antiquark annihilation $q\bar{q} \to \gamma g$ is taken.
For the quark distribution functions inside the pion and proton, we
used expressions given in Refs.~\cite{nam,watt}, respectively. We
present our results for $\sqrt s=62.4\,\, GeV$, since this value is
planned for a future Fermilab  experiment.

Obtained results are visualized in Figs. 2-15. In all figures we
represent the choice of pion distribution amplitudes
Eqs.(\ref{asy})-(\ref{HW}) by different line types: $\Phi_{asy}(x)$
as solid black line, $\Phi^{hol}(x)$ as dashed red line,
$\Phi_{VSBGL}^{hol}(x)$ as dotted blue line, $\Phi_{CZ}(x,Q^2)$ as
dash-dot magenta line, $\Phi_{BMS}(x,Q^2)$  as dash-double dot olive
line, and $\Phi_{HW}(x,Q^2)$ as short dash navy line.

First, we compare higher-twist cross sections obtained with the PMC
and BLM approaches. In Fig. 2 and 3 the dependence of higher-twist
cross sections $(\Sigma_{g}^{HT})_{PMC}$ and
$(\Sigma_{g}^{HT})_{BLM}$ are shown as a function of the gluon
transverse momentum $p_{T}$ for various choices the pion
distribution amplitudes at $y=0$. It can be observed from those
figures that the higher-twist cross section is monotonically
decreasing with the increase of the transverse momentum of the
gluon. In the region $2\,\,GeV/c<p_T<30\,\,GeV/c$ the
$(\Sigma_{g}^{HT})_{PMC}$ cross sections of the process $\pi^{+}p
\to \gamma X$ decrease in the range between $1.72\cdot10^{-2}\mu
b/GeV^{2}$ to $2.13\cdot10^{-13}\mu b/GeV^{2}$.

In Fig. 4, the dependence of the ratio
$(\Sigma_{g}^{HT})_{PMC}/(\Sigma_{g}^{HT})_{BLM}$ is displayed for
the process $\pi^{+} p \to g X$ as a function of $p_{T}$ for all
pion distribution amplitudes. We see that in the region
$5\,\,GeV/c<p_T<30\,\,GeV/c$, the PMC cross sections  for all
distribution amplitudes are enhanced by about factor of 2 relative
to corresponding BLM cross sections. We also see that the ratio
$(\Sigma_{g}^{HT})_{PMC}/(\Sigma_{g}^{HT})_{BLM}$ for
$\Phi_{HW}(x,Q^2)$ distribution amplitude is identically equivalent
to ratios for $\Phi_{asy}$, $\Phi^{hol}(x)$ and
$\Phi_{VSBGL}^{hol}(x)$ distribution amplitudes.

In Fig. 5 and 6 the dependence of higher-twist cross sections
$(\Sigma_{g}^{HT})_{PMC}$ and $(\Sigma_{g}^{HT})_{BLM}$ on the
process $\pi^{-} p \to g X$ is displayed as a function of the gluon
transverse momentum $p_{T}$ for the pion distribution amplitudes
presented in Eqs.(3.6)-(3.11) at $y=0$. As is seen in Fig. 5, in the
region $2\,\,GeV/c<p_T<30\,\,GeV/c$, PMC cross section for the
$\Phi_{HW}(x,Q^2)$ distribution amplitude is enhanced by about 2
orders of magnitude relative to the all other distribution
amplitudes. The similar dependence of the BLM cross sections is
shown in Fig. 6.

The dependence of the ratios
$(\Sigma_{g}^{HT})_{PMC}/(\Sigma_{g}^{HT})_{BLM}$,
 $(\Sigma_{g}^{HT})_{PMC}/(\Sigma_{g}^{LT})$  in the process
$\pi^{-} p \to g X$ as a function of  $p_{T}$ for the various pion
distribution amplitudes is displayed in Figs. 7-9. One of the
interesting results is that ratios
$(\Sigma_{g}^{HT})_{PMC}/(\Sigma_{g}^{HT})_{BLM}$ for processes
$\pi^{+} p \to g X$  and $\pi^{-} p \to g X$ are identically
equivalent. In Figs. 8-9 we show ratios
$(\Sigma_{g}^{HT})_{PMC}/(\Sigma_{g}^{LT})$ for processes $\pi^{\pm}
p \to g X$ as a function of  $p_{T}$. As is seen from these figures,
in the region $2\,\,GeV/c<p_T<22\,\,GeV/c$,  leading-twist cross
sections is enhanced by about 2 to 3 orders of magnitude relative to
the PMC cross sections for all pion distribution amplitudes.

In Figs. 10-15, the dependence of higher-twist cross sections
$(\Sigma_{g}^{HT})_{PMC}$, $(\Sigma_{g}^{HT})_{BLM}$, ratios
$(\Sigma_{g}^{HT})_{PMC}/(\Sigma_{g}^{HT})_{BLM}$,
 $(\Sigma_{g}^{HT})_{PMC}/(\Sigma_{g}^{LT})$ on processes
$\pi^{\pm} p \to g X$ is shown as a function of the rapidity $y$ of
the gluon at the transverse momentum of the gluon $p_T=4.9\,\,
GeV/c$. It is seen in Figs. 10-12 that the cross sections in the PMC
and BLM approaches, and the ratio for all distribution amplitudes of
pion, have a maximum approximately at the point $y=-2$. Notice that
the maximum for the CZ amplitude in PMC is more pronounced: due to
PMC the cross section for the $\Phi_{HW}(x,Q^2)$ distribution
amplitude is enhanced by about 2 to 3 orders of magnitude relative
to all other distribution amplitudes. The same dependence for the
BLM cross sections is seen in Fig. 11. In Fig. 12 we show the ratio
PMC/BLM cross sections as a function of the rapidity of gluons as
$\pi^{+}p\to gX$. We see that the ratios for all distribution
amplitudes are close to each other. Similar results for $\pi^{-} p
\to gX$ are displayed in Fig. 13-15. We think that this feature of
the comparison between PMC and BLM cross sections may help to
explain theoretical interpretations with future experimental data
for the inclusive gluon production cross section in the pion-proton
collisions. Higher-twist cross section obtained in our study should
be observable at hadron collider.

We can see in Figs. 2, 3, 5, 6, 10, 11, 13, and 14 that cross
sections for twist-3 distribution amplitudes calculated within
PMC/BLM schemes are significantly different from the others. This
behaviour can be explained  by the following: first, in the twist-3
distribution amplitude defined in Eq.(3.11), the usual helicity
components $(\lambda_1+\lambda_2=0)$ have been taken into account.
However, the higher helicity components $(\lambda_1+\lambda_2=\pm1)$
that come from the spin-space Wigner rotation have not been
considered. Second, in the twist-3 distribution amplitude
construction, the contribution of the intrinsic quark propagator
$k_T$ is also included. Directly we can obtain, that the quark
propagator,  gives essential contribution for the twist-3
distribution amplitude,  depending  on the transverse
momentum~\cite{Huang}.

As one can see from the figures, there is a  large difference
between the cross sections calculated by PMC and BLM schemes. Main
reasons for this are the following: as we know, the PMC scheme is
defined in the form $(\mu_{1}^{2})^{PMC}= (1-x)\hat se^{-5/3}$  and
$(\mu_{2}^{2})^{PMC}= -x\hat ue^{-5/3}$,  but the BLM scale in the
form $(\mu_{1}^{2})^{BLM}= (1-x)\hat s$  and $(\mu_{2}^{2})^{PMC}=
-x\hat u$. PMC and BLM schemes differ with the factor $e^{-5/3}$ and
also take into account the running coupling constant proportional to
$1/ln(\mu^2/\Lambda^2)$. Therefore, cross sections calculated within
the PMC scale are enhanced by cross section calculated within the
BLM scale.

Notice that in Figs. 4, 7, 12, 15, the curves for ratio
$(\Sigma_{g}^{HT})_{PMC}/(\Sigma_{g}^{HT})_{BLM}$ [as shown in
figures: $asy(PMC)/asy(BLM)$, $hol(PMC)/hol(BLM)$,
$VSBGL(PMC)/VSBGL(BLM)$, $HW(PMC)/HW(BLM))$] for distribution
amplitudes $\Phi^{hol}(x)$, $\Phi_{asy}(x)$, $\Phi_{VSBGL}^{hol}(x)$
and $\Phi_{HW}(x,Q^2)$ cannot be resolved from each other because
the corresponding cross section ratios are almost exactly equal.

\section{CONCLUSIONS}
\label{conc}

In this paper, we have presented the inclusive gluon production
cross section in the process $\pi p$ collisions via higher-twist
mechanisms within holographic and perturbative QCD. For calculation
of the cross section, the renormaliation scale in running coupling
is applied within PMC and BLM approaches. The results obtained from
the approaches PMC and BLM are different in some regions. These
results also depend on the form of the pion distribution amplitudes.
We compared BLM and PMC cross sections of the inclusive gluon
production in the processes $\pi^{-} p \to g X$ and $\pi^{+} p \to g
X$. Our results show in both cases that the inclusive gluon
production cross section in the process $\pi^{-} p \to g X$ is
suppress over the inclusive gluon production cross section in the
process $\pi^{+} p \to g X$. Notice that the inclusive gluon
production spectrum can be measured with large precision, so results
obtained in this study will help further tests for hadron dynamics
at large $p_{T}$. It is also observed that higher-twist cross
sections  are proportional to the third order of $\alpha_{s}(Q^2)$,
but the leading-twist is linearly proportional to $\alpha_{s}(Q^2)$.
Therefore, their ratio strongly depends of the
$\alpha_{s}^{2}(Q^2)$.

Our calculation show that the higher helicity components in the
higher-twist distribution amplitude make significant contributions
to the cross section. We also note that the intrinsic transverse
momentum of the distribution  amplitude is very important for the
cross section, and without including these effects one overestimates
the results.

In our opinion higher-twist cross section  obtained by PMC approach
are approximately equivalent to the resummed cross section in the
same process. Also, we hope that the PMC approach can be applied to
a wide class of perturbatively calculable collider and  other
processes. Results obtaining within the PMC approach can improve the
precision of tests of the Standard Model and enhance sensitivity to
new physical phenomena.

\section*{Acknowledgments}

The authors thank to Stanley J. Brodsky for useful discussions. We
are also grateful to  Dr. S.H.Aydin and A.Mustafayev for carefully
reading the paper and providing useful comments. This work is
supported by TUBITAK under grant number 2221(Turkey). A. I. A
acknowledges hospitality the Department of Physics of Karadeniz
Technical University during this work. The author C.A thanks to
member of the Department of Physics at University of Surrey
especially to Dr. Paul Stevenson giving opportunity to research in
their institute.

\newpage

\begin{figure}[!hbt]
\vskip -0.5cm \epsfxsize 15cm \centerline{\epsfbox{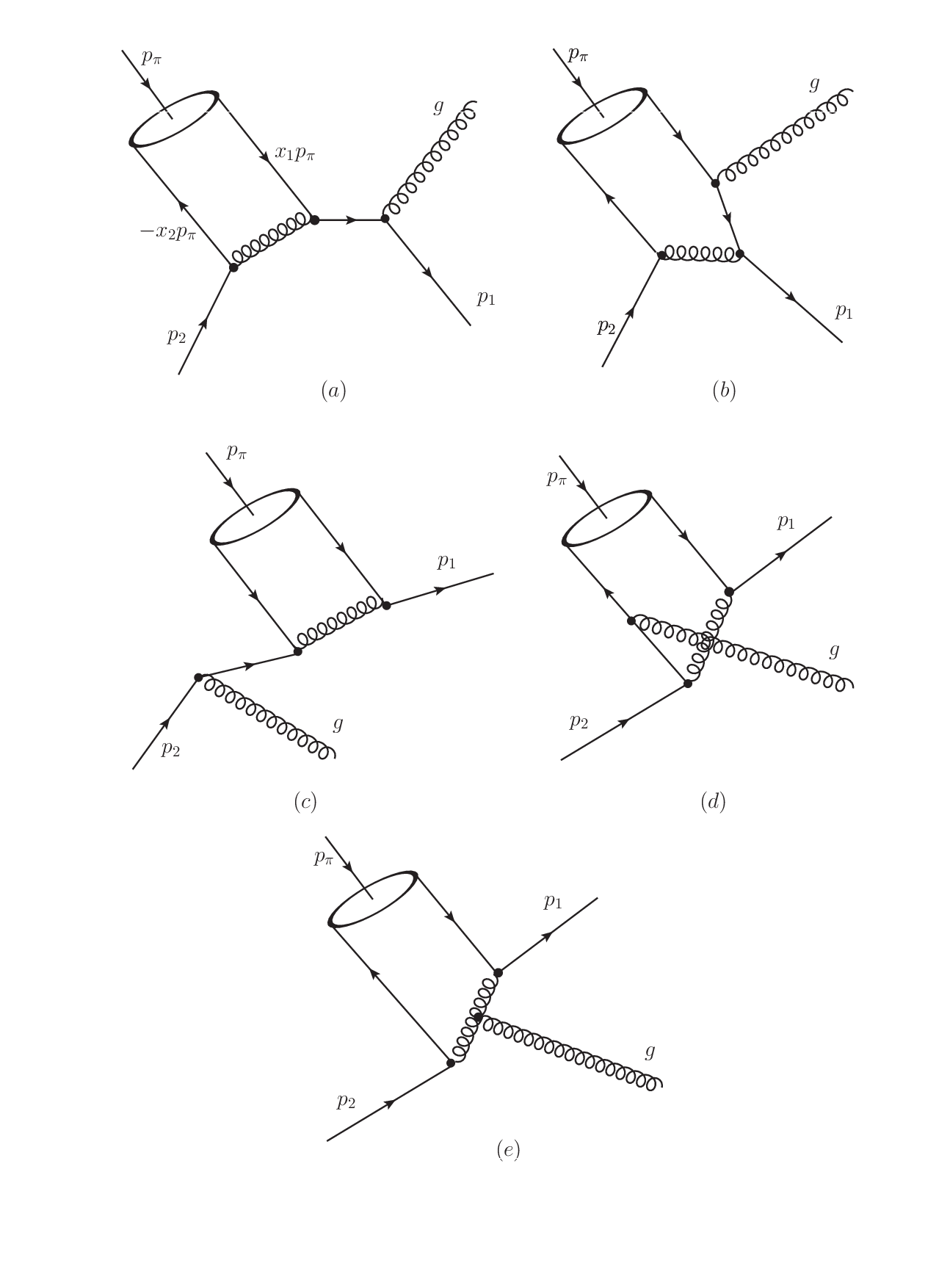}} \vskip
-2cm \caption{ Full set of QCD Feynman diagrams for higher-twist
subprocess $\pi q\to g q$.} \label{Fig1}
\end{figure}

\begin{figure}[!hbt]
\vskip 1.2cm\epsfxsize 11.8cm \centerline{\epsfbox{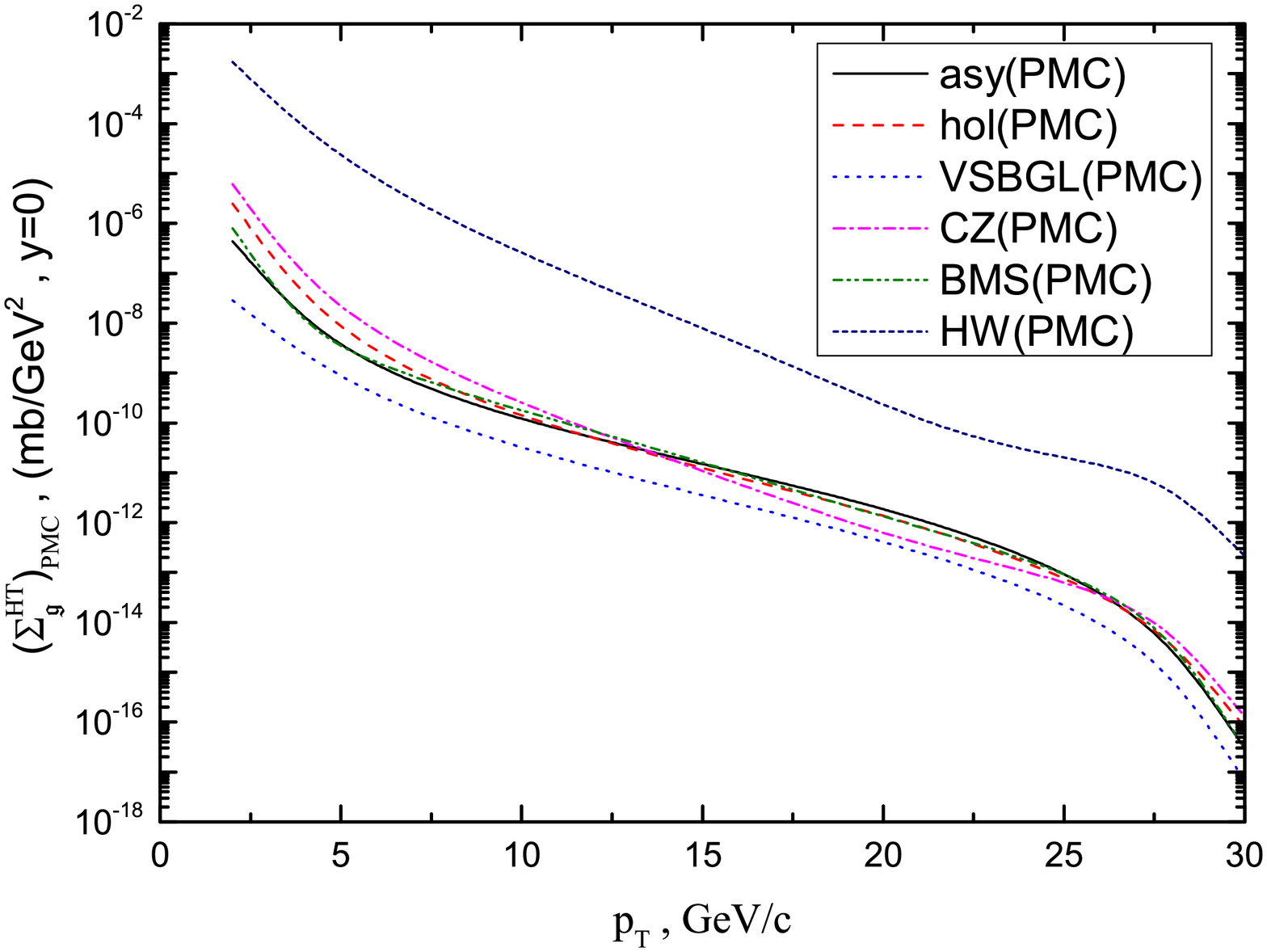}}
\vskip-0.2cm \caption{Higher-twist $\pi^{+} p\to g X$ inclusive
gluon production cross section $(\Sigma_{g}^{HT})_{PMC}$ as a
function of the $p_{T}$ of the gluon at $\sqrt s=62.4\,\, GeV$. }
\label{Fig2}
\end{figure}

\begin{figure}[!hbt]
\vskip 1.2cm \epsfxsize 11.8cm \centerline{\epsfbox{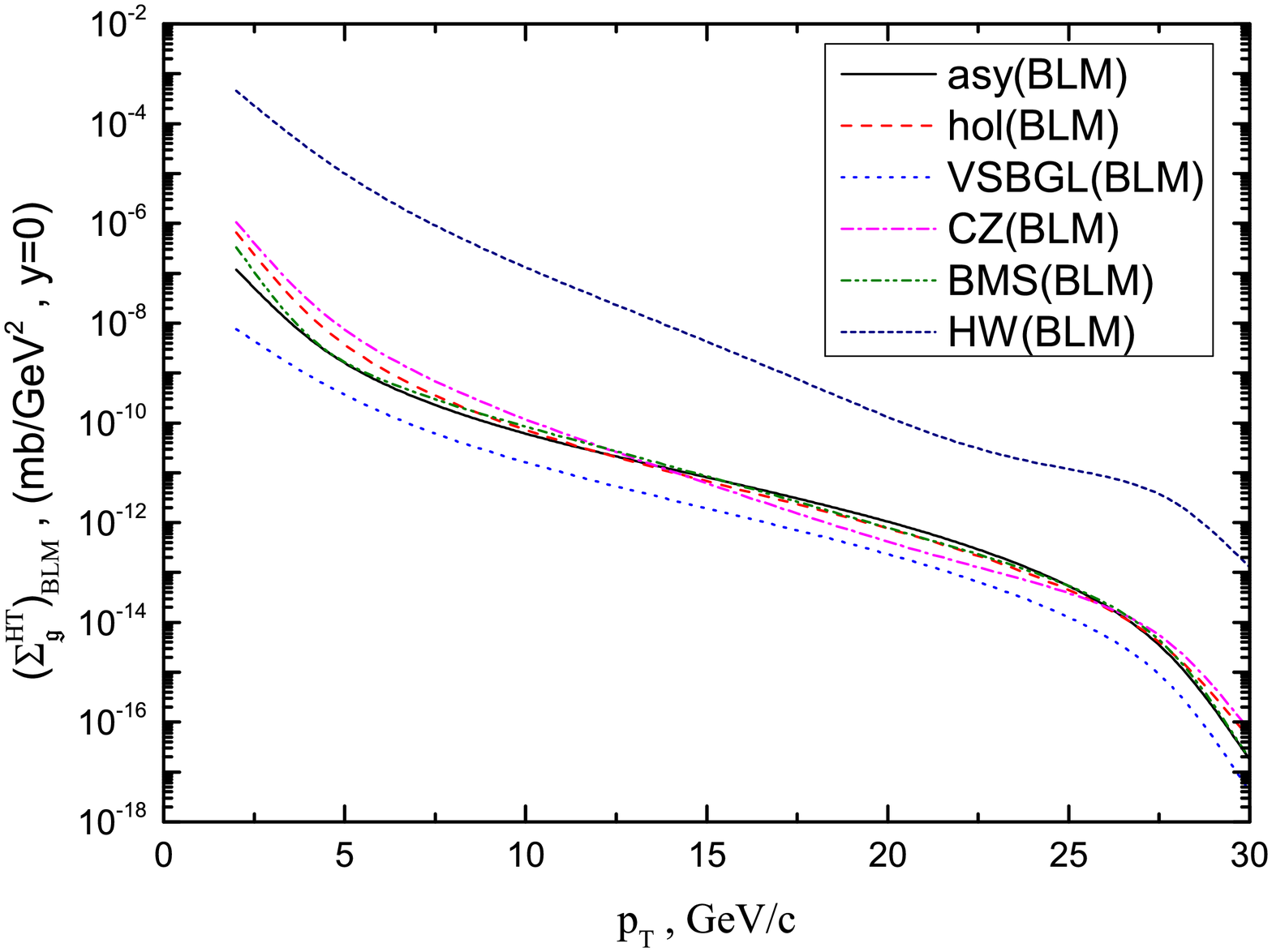}}
\vskip-0.2cm \caption{Higher-twist $\pi^{+} p\to g X$ inclusive
gluon production cross section $(\Sigma_{g}^{HT})_{BLM}$ as a
function of the $p_{T}$ of the gluon at $\sqrt s=62.4\,\, GeV$.}
\label{Fig3}
\end{figure}

\begin{figure}[!hbt]
\vskip -1.2cm\epsfxsize 11.8cm \centerline{\epsfbox{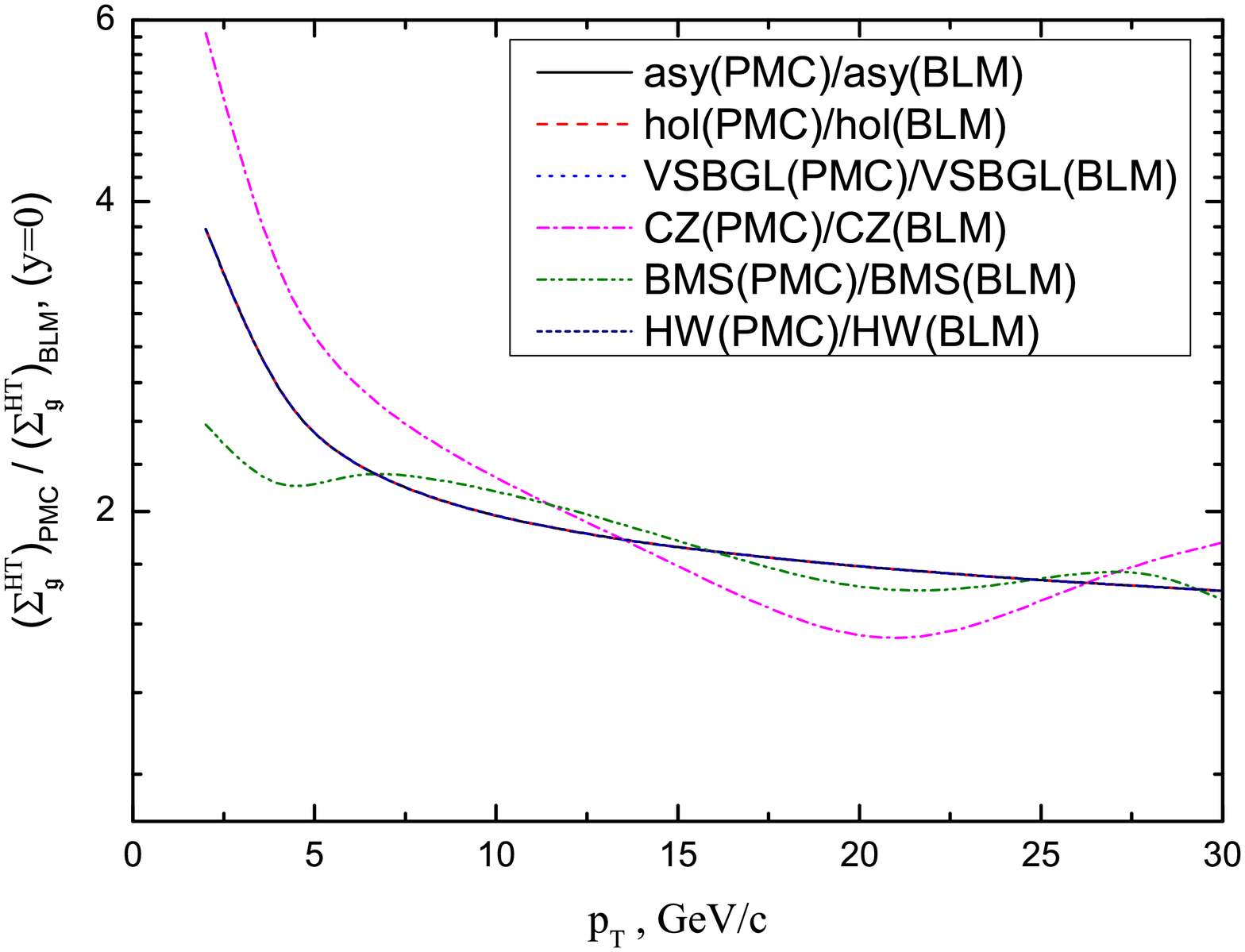}}
\vskip-0.2cm \caption{Ratio
$(\Sigma_{g}^{HT})_{PMC}/(\Sigma_{g}^{HT})_{BLM}$, in the process
$\pi^{+} p\to g X$, where higher-twist contribution are calculated
for the gluon rapidity $y=0$ at $\sqrt s=62.4\,\, GeV$ as a function
of the $p_{T}$ of the gluon. Notice that curves for
asy(PMC)/asy(BLM), hol(PMC)/hol(BLM), VSBGL(PMC)/VSBGL(BLM),
HW(PMC)/HW(BLM) pion distribution amplitudes completely overlap.}
\label{Fig4}
\end{figure}

\begin{figure}[!hbt]
\vskip 1.2cm\epsfxsize 11.8cm \centerline{\epsfbox{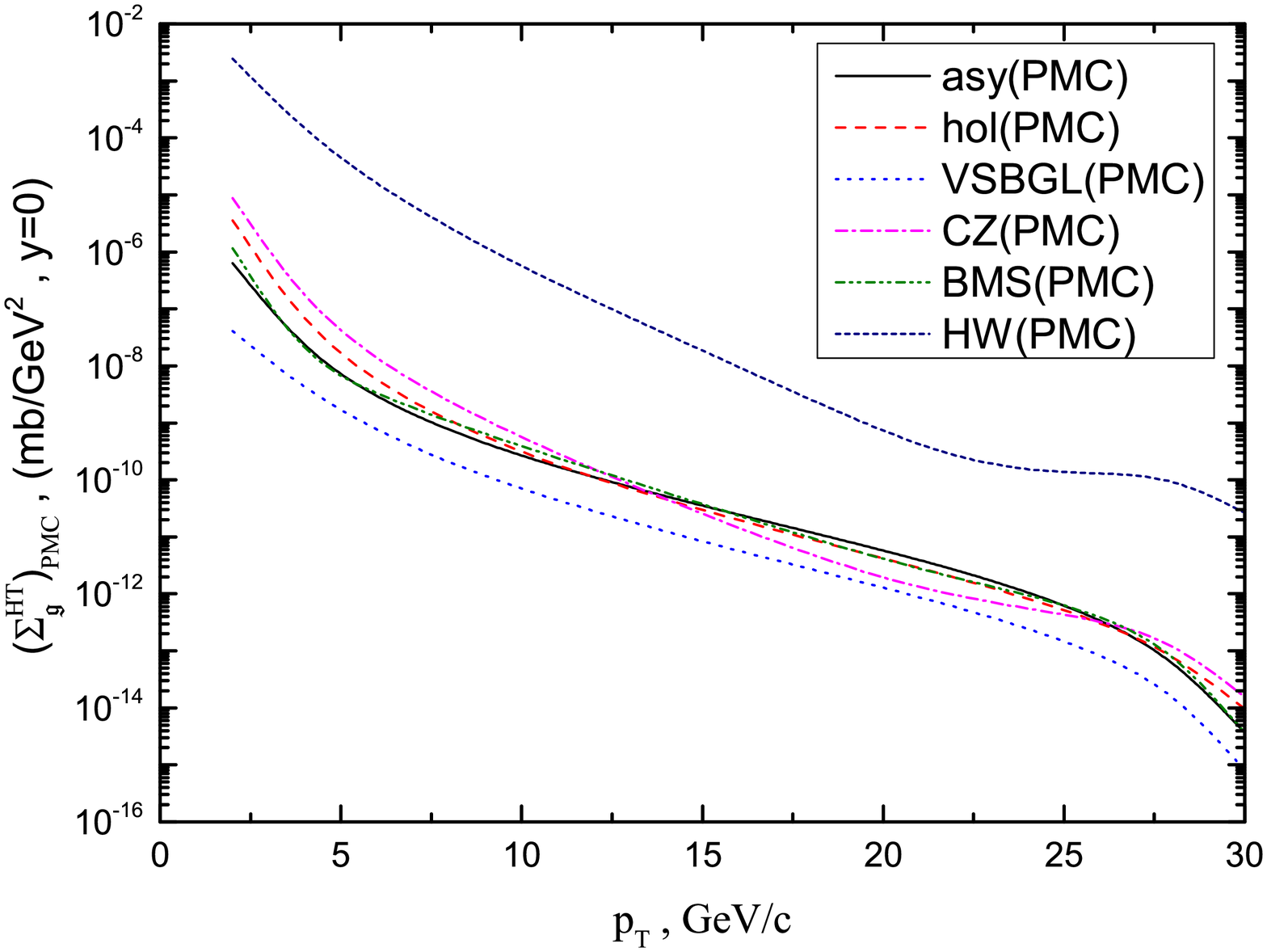}}
\vskip-0.2cm \caption{Higher-twist $\pi^{-} p\to g X$ inclusive
gluon production cross section $(\Sigma_{g}^{HT})_{PMC}$ as a
function of the $p_{T}$ of the gluon at $\sqrt s=62.4\,\, GeV$.}
\label{Fig5}
\end{figure}

\begin{figure}[!hbt]
\vskip -1.2cm\epsfxsize 11.8cm \centerline{\epsfbox{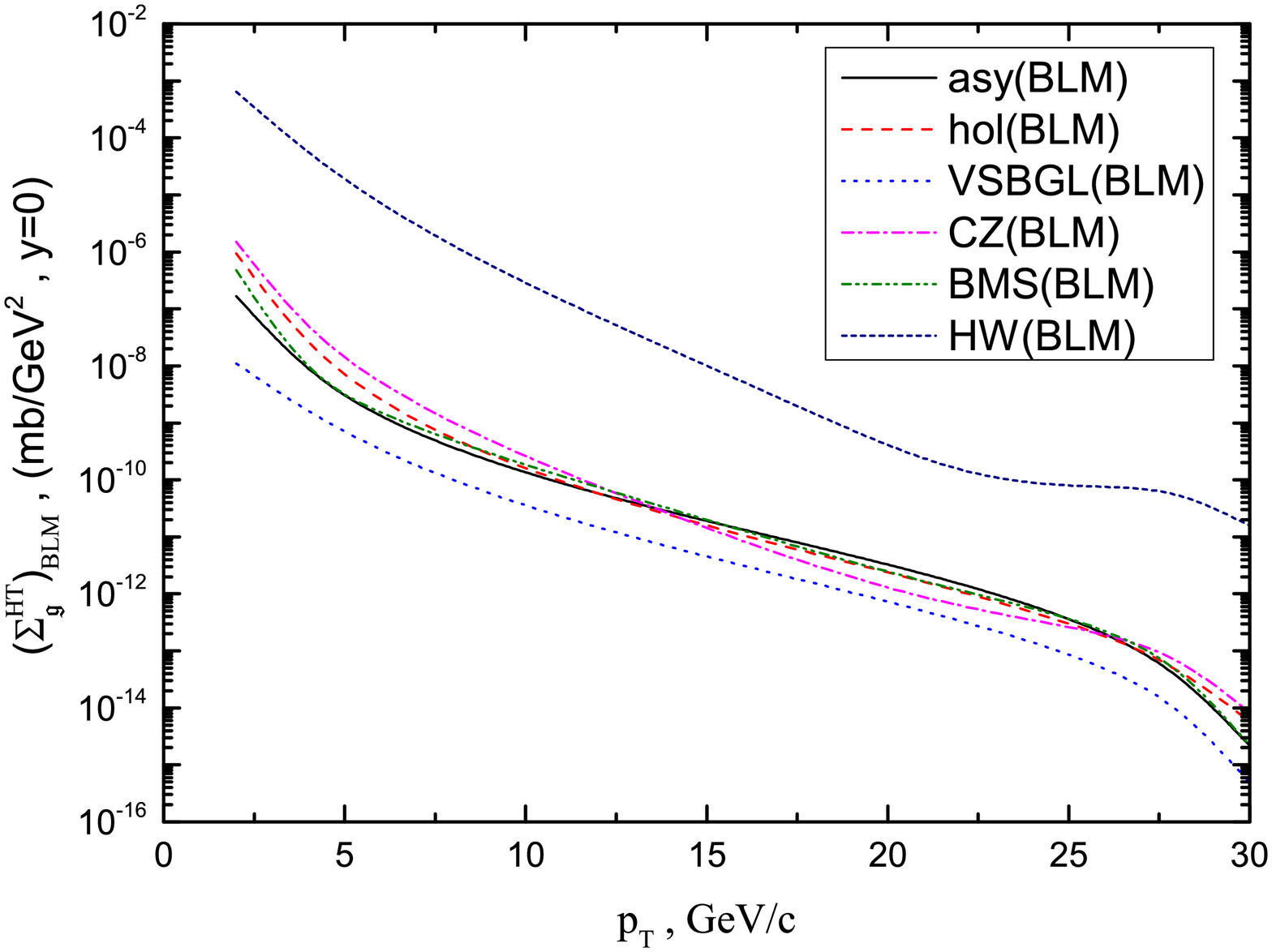}}
\vskip-0.2cm \caption{Higher-twist $\pi^{-} p\to g X$ inclusive
gluon production cross section $(\Sigma_{g}^{HT})_{BLM}$ as a
function of the $p_{T}$ of the gluon at $\sqrt s=62.4\,\, GeV$.}
\label{Fig6}
\end{figure}

\begin{figure}[!hbt]
\vskip 1.2cm\epsfxsize 11.8cm \centerline{\epsfbox{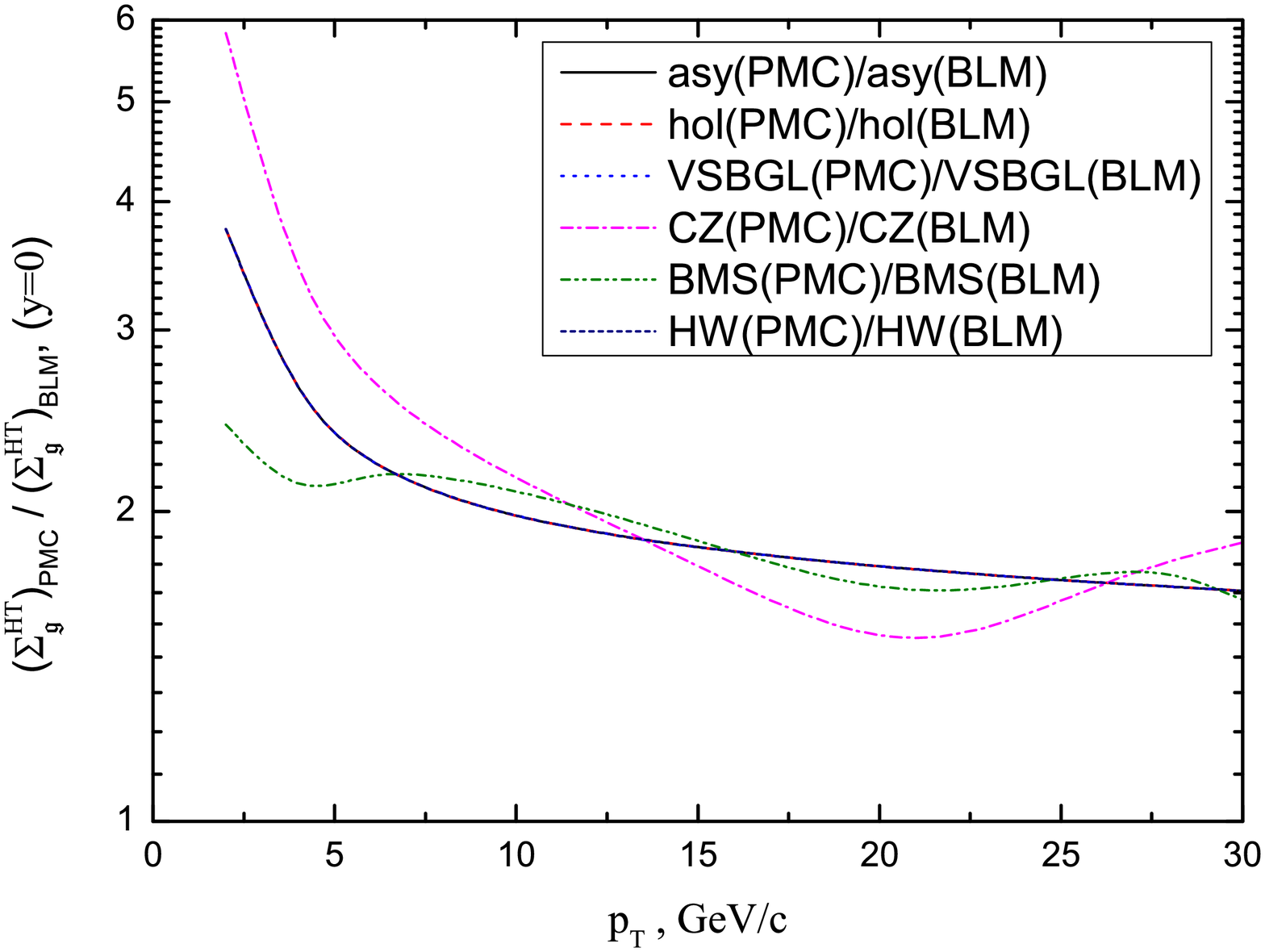}}
\vskip-0.2cm \caption{Ratio
$(\Sigma_{g}^{HT})_{PMC}/(\Sigma_{g}^{HT})_{BLM}$, in the process
$\pi^{-} p\to g X$, where higher-twist contribution are calculated
for the gluon rapidity $y=0$ at $\sqrt s=62.4\,\, GeV$ as a function
of the $p_{T}$ of the gluon. Notice that curves for
asy(PMC)/asy(BLM), hol(PMC)/hol(BLM), VSBGL(PMC)/VSBGL(BLM),
HW(PMC)/HW(BLM) pion distribution amplitudes completely overlap.}
\label{Fig7}
\end{figure}

\begin{figure}[!hbt]
\vskip-1.2cm \epsfxsize 11.8cm \centerline{\epsfbox{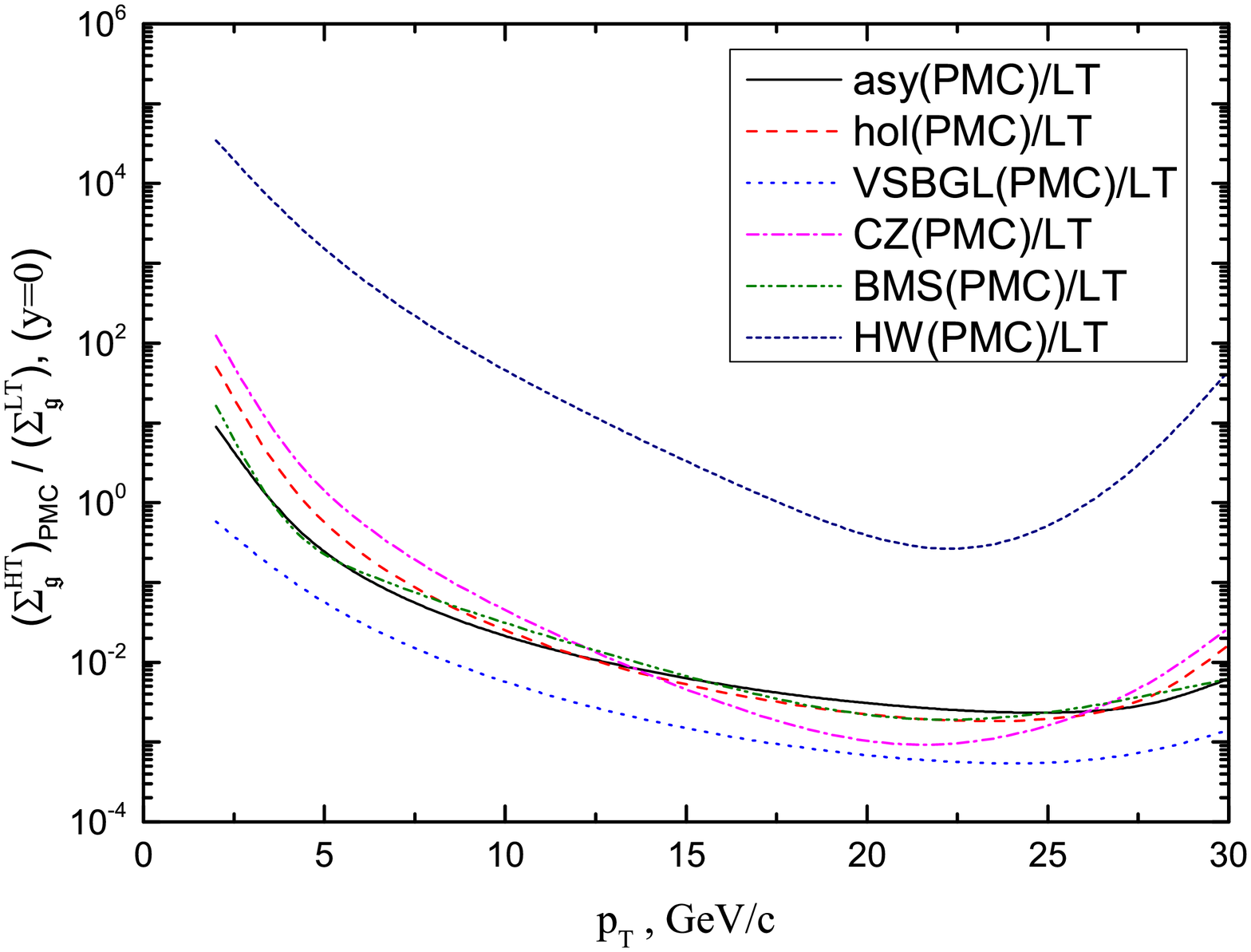}}
\vskip-0.2cm \caption{Ratio
$(\Sigma_{g}^{HT})_{PMC}/(\Sigma_{g}^{LT})$, in the process $\pi^{+}
p\to g X$, as a function of the $p_{T}$ of the gluon at  $\sqrt
s=62.4\,\, GeV$.} \label{Fig8}
\end{figure}

\begin{figure}[!hbt]
\vskip 0.8cm \epsfxsize 11.8cm \centerline{\epsfbox{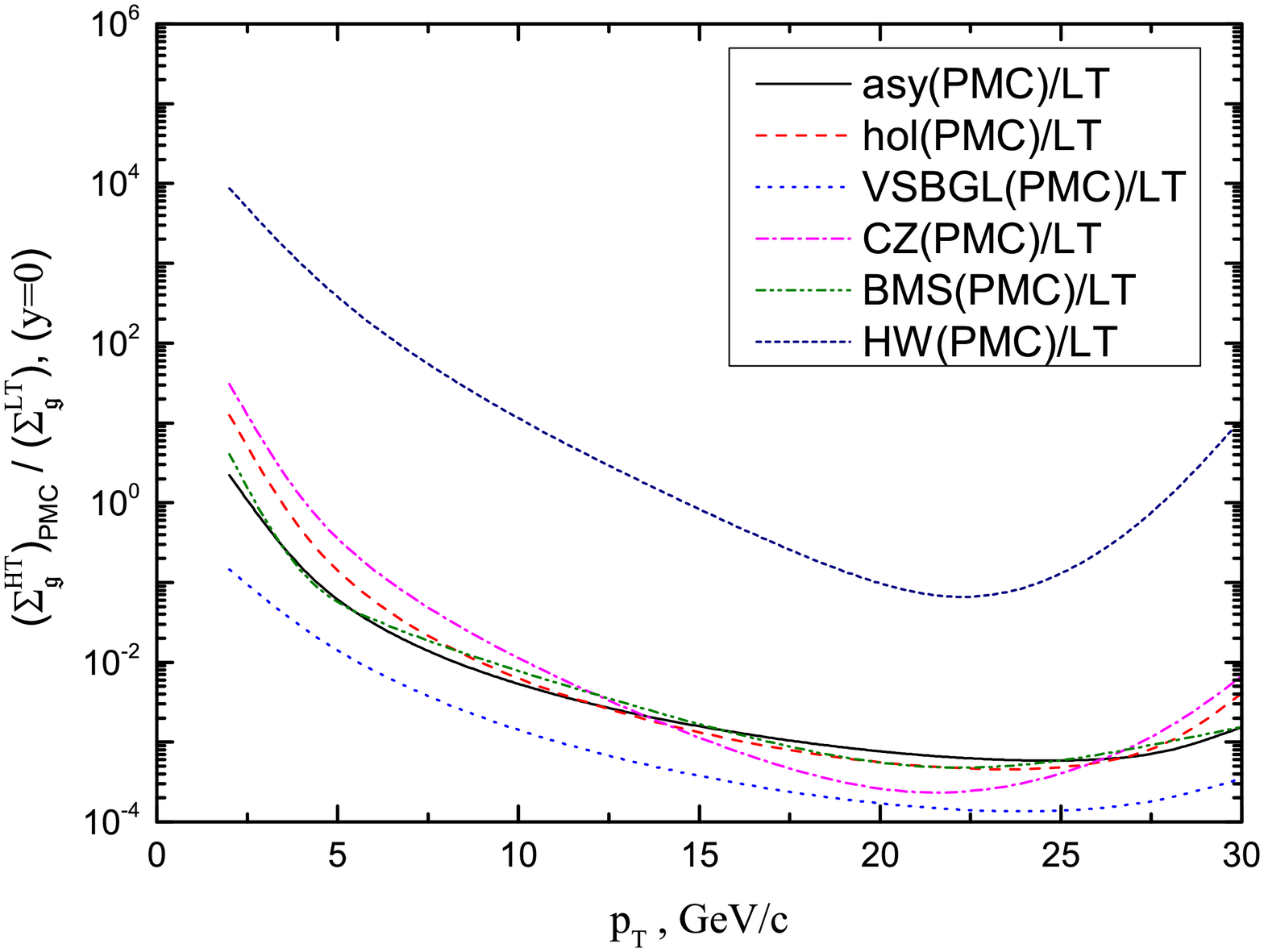}}
\vskip-0.2cm \caption{Ratio
$(\Sigma_{g}^{HT})_{PMC}/(\Sigma_{g}^{LT})$, in the process $\pi^{-}
p\to g X$, as a function of the $p_{T}$ of the gluon at $\sqrt
s=62.4\,\, GeV$.} \label{Fig9}
\end{figure}

\begin{figure}[!hbt]
\vskip-1.2cm \epsfxsize 11.8cm \centerline{\epsfbox{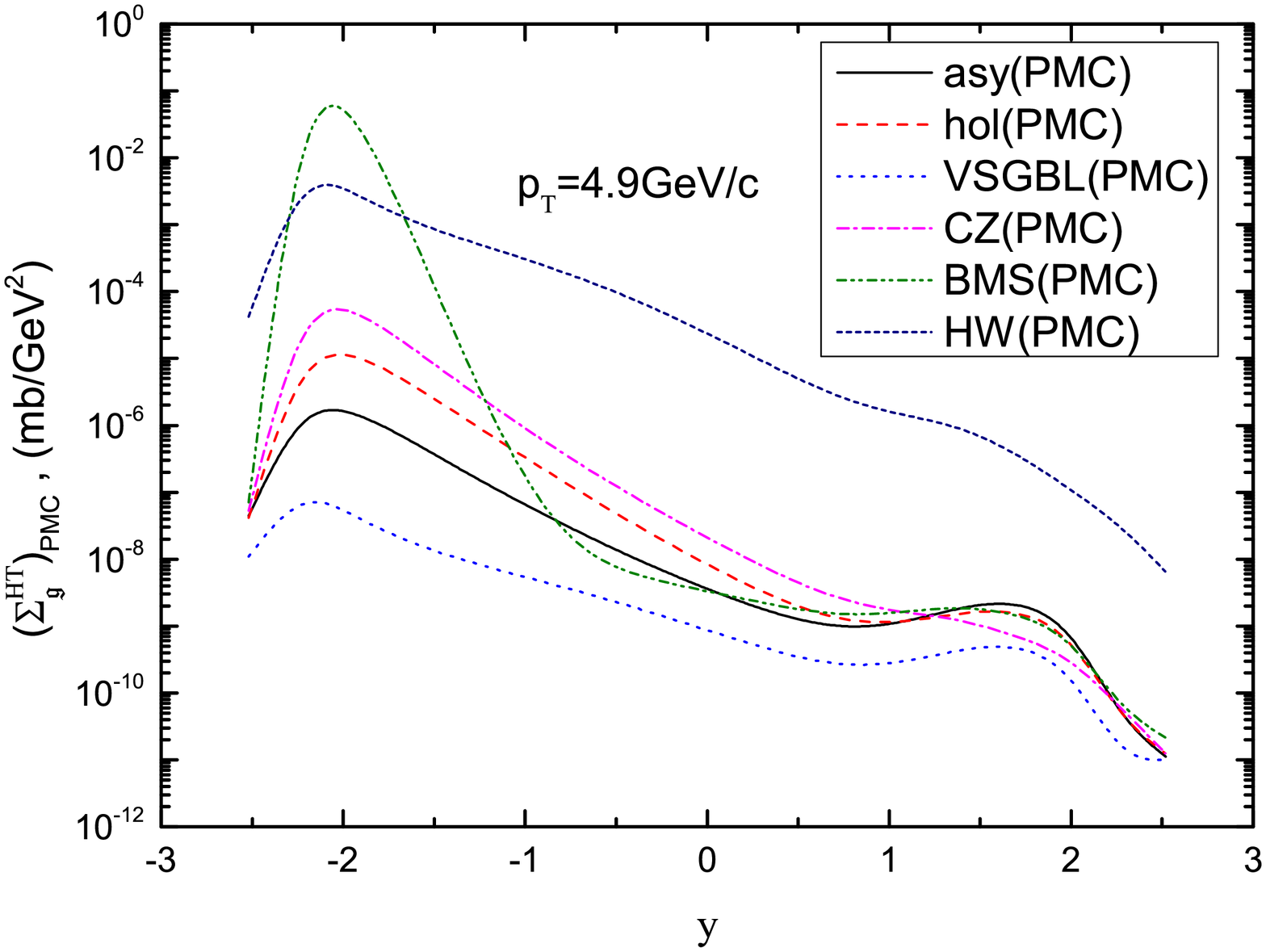}}
\vskip-0.2cm \caption{Higher-twist $\pi^{+} p\to g X$ inclusive
gluon production cross section $(\Sigma_{g}^{HT})_{PMC}$, as a
function of the $y$ rapidity of the gluon at $p_T=4.9\,\, GeV/c$, at
$\sqrt s=62.4\,\, GeV$.} \label{Fig10}
\end{figure}

\begin{figure}[!hbt]
\vskip-1.2cm\epsfxsize 11.8cm \centerline{\epsfbox{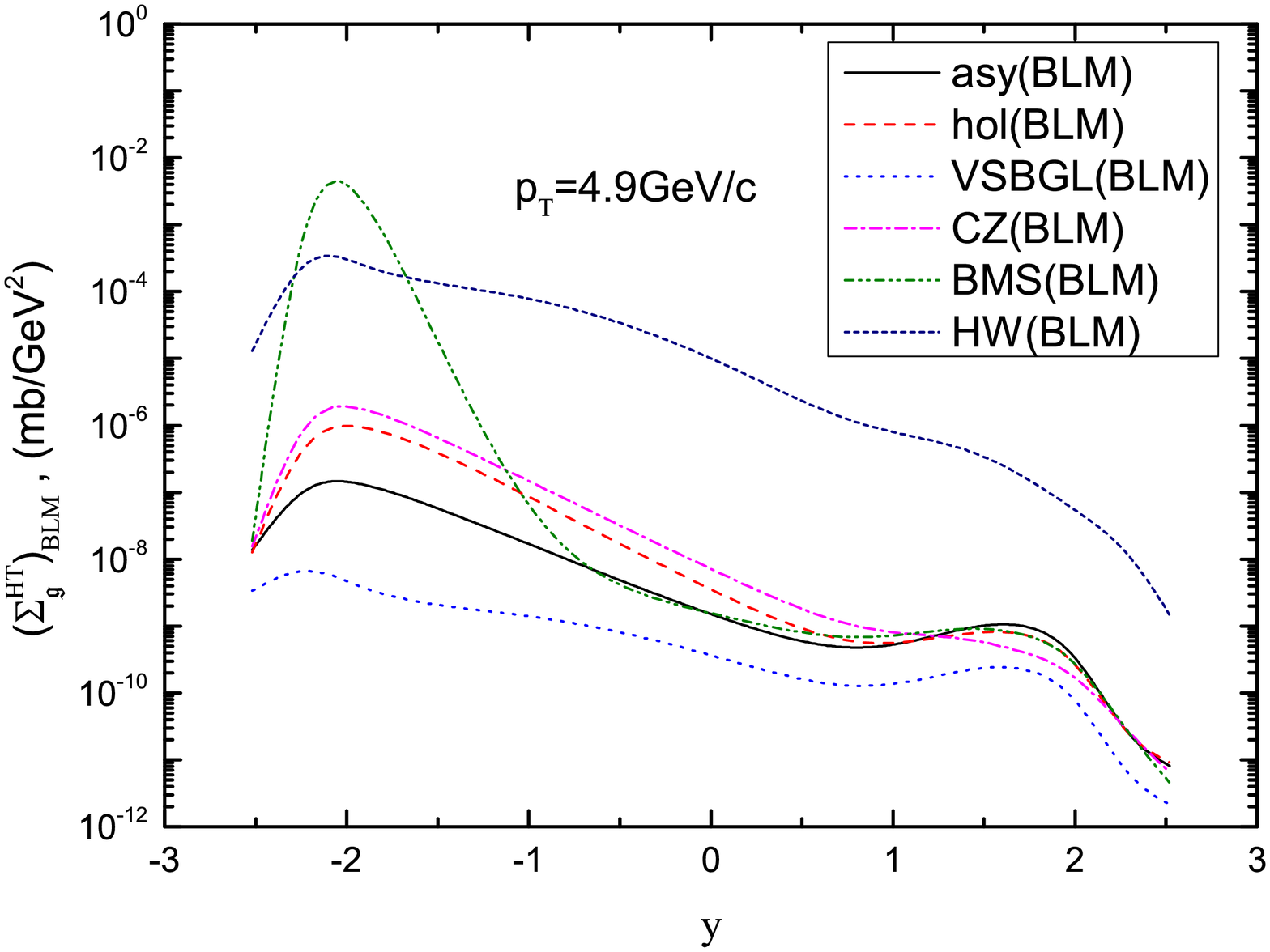}}
\vskip-0.2cm \caption{Higher-twist $\pi^{+} p\to g X$ inclusive
gluon production cross section $(\Sigma_{g}^{HT})_{BLM}$ , as a
function of the $y$ rapidity of the gluon at $p_T=4.9\,\, GeV/c$, at
$\sqrt s=62.4\,\, GeV$.} \label{Fig11}
\end{figure}

\begin{figure}[!hbt]
\vskip-1.2cm\epsfxsize 11.8cm \centerline{\epsfbox{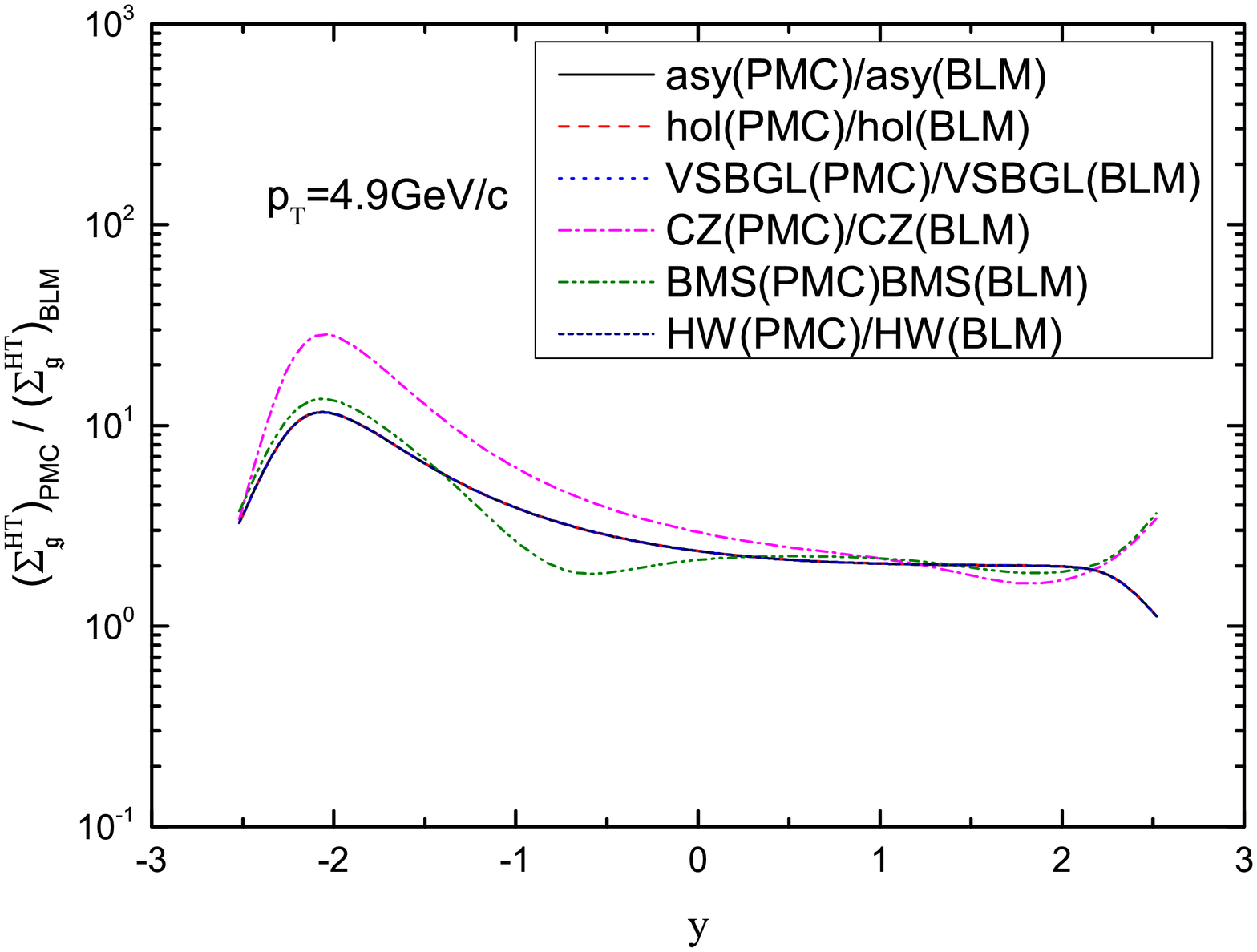}}
\vskip-0.2cm \caption{Ratio
$(\Sigma_{g}^{HT})_{PMC}/(\Sigma_{g}^{HT})_{BLM}$, in the process
$\pi^{+} p\to g X$, as a function of the $y$ rapidity of the gluon
at $p_T=4.9\,\, GeV/c$, at $\sqrt s=62.4\,\, GeV$. Notice that
curves for asy(PMC)/asy(BLM), hol(PMC)/hol(BLM),
VSBGL(PMC)/VSBGL(BLM), HW(PMC)/HW(BLM) pion distribution amplitudes
completely overlap.} \label{Fig12}
\end{figure}

\begin{figure}[!hbt]
\vskip-1.2cm\epsfxsize 11.8cm \centerline{\epsfbox{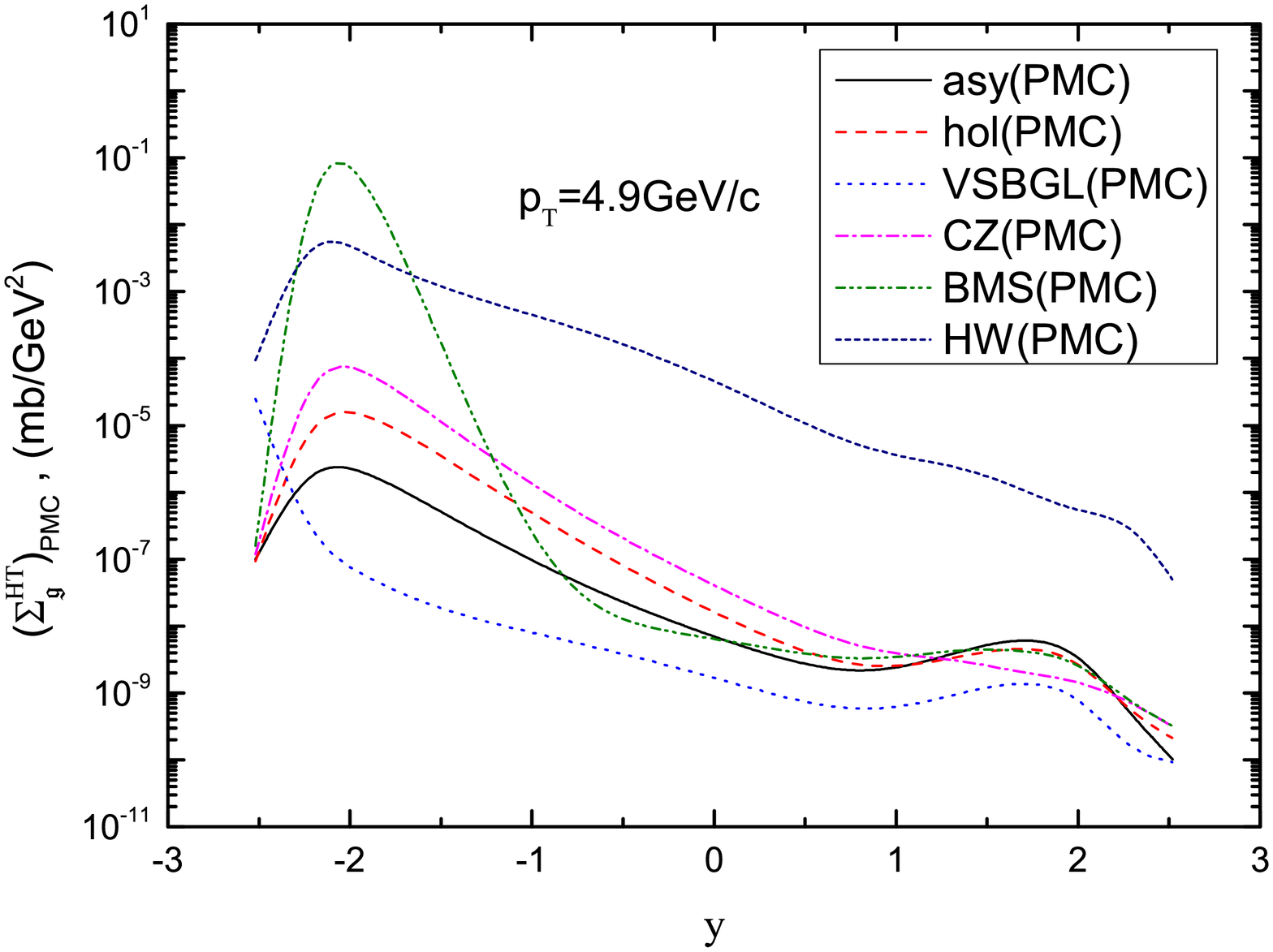}}
\vskip-0.2cm \caption{Higher-twist $\pi^{-} p\to g X$ inclusive
gluon production cross section $(\Sigma_{g}^{HT})_{PMC}$, as a
function of the $y$ rapidity of the gluon at $p_T=4.9\,\, GeV/c$, at
$\sqrt s=62.4\,\, GeV$.} \label{Fig13}
\end{figure}

\begin{figure}[!hbt]
\vskip -1.2cm \epsfxsize 11.8cm \centerline{\epsfbox{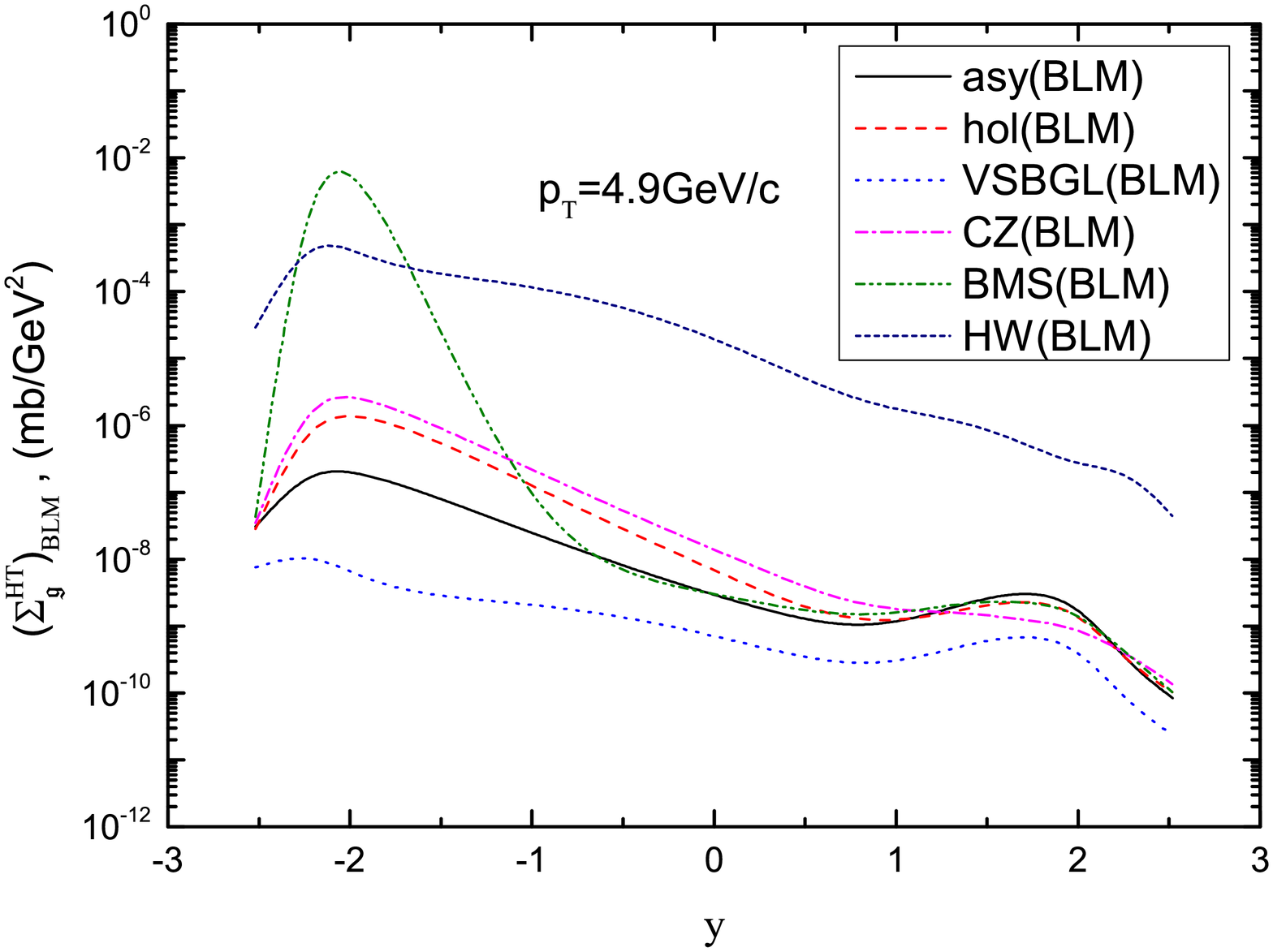}}
\vskip-0.2cm \caption{Higher-twist $\pi^{-} p\to g X$ inclusive
gluon production cross section $(\Sigma_{g}^{HT})_{BLM}$, as a
function of the $y$ rapidity of the gluon at $p_T=4.9\,\, GeV/c$, at
$\sqrt s=62.4\,\, GeV$.} \label{Fig14}
\end{figure}

\begin{figure}[!hbt]
\vskip -1.2cm \epsfxsize 11.8cm \centerline{\epsfbox{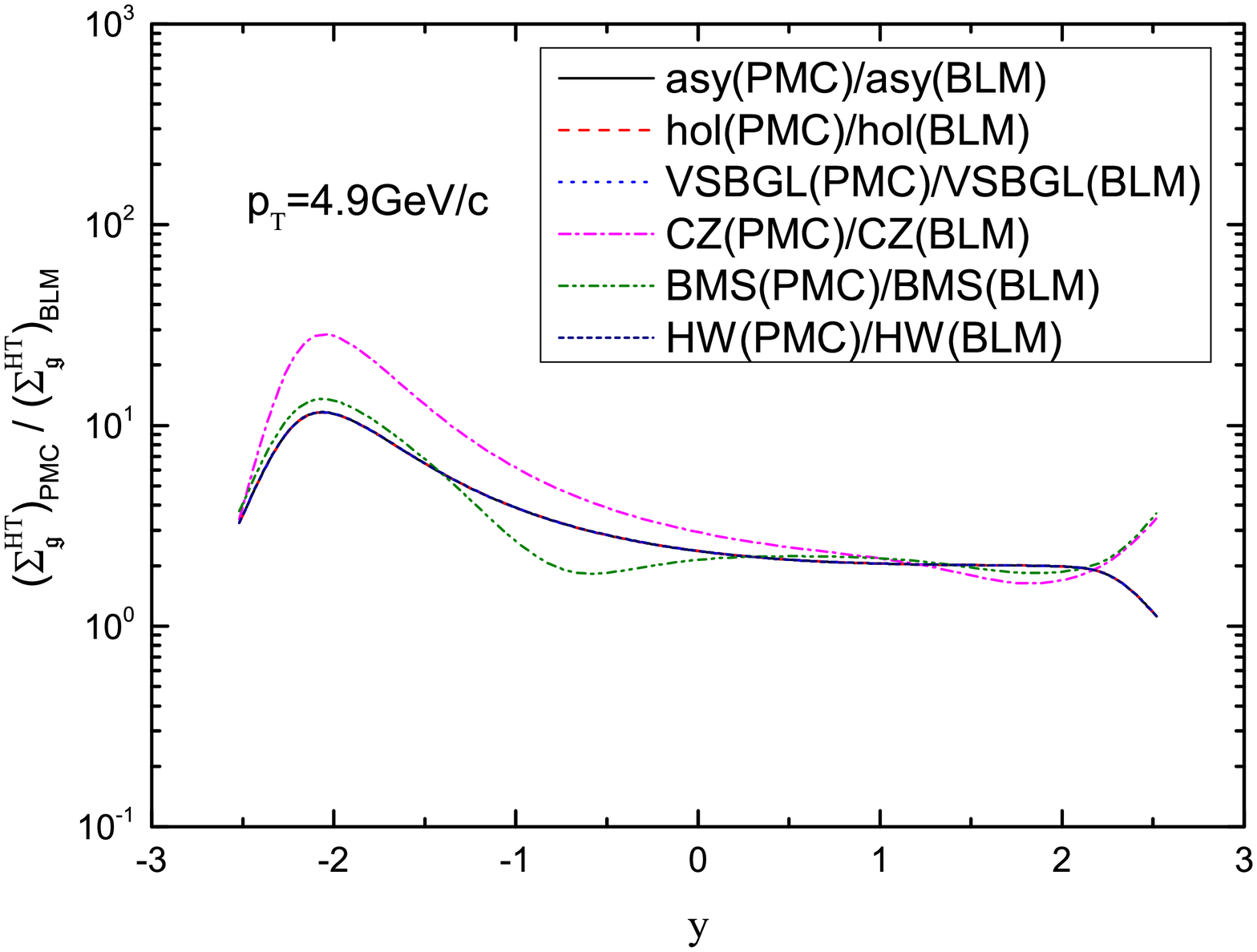}}
\vskip-0.2cm \caption{Ratio
$(\Sigma_{g}^{HT})_{PMC}/(\Sigma_{g}^{HT})_{BLM}$, in the process
$\pi^{-} p\to g X$, as a function of the $y$ rapidity of the gluon
at $p_T=4.9\,\, GeV/c$, at $\sqrt s=62.4\,\, GeV$. Notice that
curves for asy(PMC)/asy(BLM), hol(PMC)/hol(BLM),
VSBGL(PMC)/VSBGL(BLM), HW(PMC)/HW(BLM) pion distribution amplitudes
completely overlap.} \label{Fig15}
\end{figure}


\begin{thebibliography}{99}
\bibitem{Brodsky1}
 S. J. Brodsky, G. L.~Lepage, and P. B.~Mackenize, Phys. Rev. D\textbf{
28}, 228 (1983).
\bibitem{Brodsky10}
S. J. Brodsky and  L. Di Glustino, Phys. Rev. D\textbf{ 86}, 085026
(2012).
\bibitem{Bagger}
J. A.~Bagger and J. F.~Gunion, Phys. Rev. D\textbf{ 29}, 40 (1984).
\bibitem{Bagger1}
A.~Bagger and J. F.~Gunion, Phys. Rev. D\textbf{ 25}, 2287 (1982).
\bibitem{Baier}
V. N.~Baier and A. Grozin, Phys. Lett. \textbf{96B}, 181 (1980);
S.~Gupta, Phys. Rev. D\textbf{24}, 1169 (1981).
\bibitem{Sadykhov}
F. S.~Sadykhov and A. I.~Akhmedov, Russ.  Phys. J. \textbf{38}, 513
(1995).
\bibitem{Ahmadov1}
A. I.~Ahmadov, I. Boztosun, R. Kh. Muradov, A. Soylu, and E. A.
 Dadashov, Int. J. Mod. Phys. E\textbf{15}, 1209 (2006).
\bibitem{Ahmadov2}
A. I.~Ahmadov, I. Boztosun, A. Soylu, and E. A. Dadashov, Int. J.
Mod. Phys. E\textbf{17}, 1041 (2008).
\bibitem{Ahmadov3}
A. I.~Ahmadov, Coskun ~Aydin, Sh. M.~Nagiyev, A.~Hakan Yilmaz, and
E. A.~Dadashov, Phys. Rev. D\textbf{80}, 016003 (2009).
\bibitem{Ahmadov4}
A. I.~Ahmadov, Coskun ~Aydin, E. A.~Dadashov, and Sh. M.~Nagiyev,
Phys. Rev. D\textbf{81}, 054016 (2010).
\bibitem{Ahmadov5}
A. I.~Ahmadov, R. M.~Burjaliyev, Int. J. Mod. Phys. E\textbf{20},
1243 (2011).
\bibitem{Ahmadov6}
A. I.~Ahmadov, Sh. M.~Nagiyev, and
 E. A.~Dadashov , Int. J. Mod. Phys. E\textbf{21},
1250014 (2012).
 \bibitem{Ahmadov7}
A. I.~Ahmadov, C.~Aydin, and F. Keskin, Phys. Rev. D\textbf{85},
034009 (2012).
 \bibitem{Ahmadov8}
 A. I.~Ahmadov, C.~Aydin, and F. Keskin,
 Ann. Phys. \textbf{327}, 1472 (2012).
\bibitem{Ahmadov9}
A. I.~Ahmadov, C.~Aydin, and O. Uzun,  Phys. Rev. D\textbf{87},
014006 (2013).
\bibitem{Brodsky5}
S. J. Brodsky and Xing-Gang Wu, Phys. Rev. D\textbf{86}, 054018
(2012).
\bibitem{Brodsky6}
S. J. Brodsky  and Xing-Gang Wu, Phys. Rev. D\textbf{85}, 114040
(2012).
\bibitem{Brodsky7}
S. J. Brodsky and Xing-Gang Wu, Phys. Rev. D\textbf{86}, 014021
(2012).
\bibitem{Brodsky8}
S. J. Brodsky and  Xing-Gang Wu, Phys. Rev. Lett. \textbf{109},
042002 (2012).
\bibitem{Brodsky9}
S. J. Brodsky and Xing-Gang Wu, Phys. Rev. D\textbf{85}, 034038
(2012).
\bibitem{Lu}
S.J. Brodsky, Report No. SLAC-PUB-6304, 1993; S.J. Brodsky and H.J.
Lu, Report No. SLAC-PUB-6000.
\bibitem{Maitre}
 D. Maitre et al., PoS Sci., EPS-HEP(2009) 367.
 \bibitem{Bardeen}
 W. A. Bardeen, A. J. Buras, D. W. Duke, and T. Muta, Phys. Rev. D\textbf{18}, 3998 (1978).
\bibitem{Lepage2}
G. L.~Lepage and S. J.~Brodsky, Phys. Rev. D\textbf{22}, 2157
(1980).
\bibitem{Owens}
J.F. Owens, Rev. Modern Phys. \textbf{59}, 465 (1987).
\bibitem {Berger}
E. L.~Berger and S. J. Brodsky, Phys. Rev. DE\textbf{24}, 2428
(1981).
\bibitem {Lepage1}
G. P.~Lepage and S. J.~Brodsky, Phys. Lett. \textbf{87B}, 359
(1979).
\bibitem {Vega}
A. ~Vega, I. Schmidt, T. Branz, T.Gutsche, V. Lyubovitskij, Phys.
Rev. D\textbf{80}, 055014 (2009).
\bibitem {Brodsky3}
S. J. Brodsky and G. F. de Teramond, Phys. Rev. D\textbf{77}, 056007
(2008).
\bibitem {Brodsky2}
S. J.~Brodsky, Proc. Sci., LHC07 (2007) 002.
\bibitem {chernyak}
V. L.~Chernyak and A. R.~Zhitnitsky, Phys. Rep. \textbf{112}, 173
(1984).
\bibitem{Bakulev}
 A. P.~Bakulev, S. V.~Mikhailov and N. G.~Stefanis, Phys. Lett,
\textbf{B 578}, 91 (2004); A. P. Bakulev, S.V. Mikhailov, A.V.
Pimikov, and N. G. Stefanis, Phys. Rev. D\textbf{86}, 031501 (2012).
\bibitem{Huang}
T. ~Huang and X.-G. Wu, Phys. Rev. D\textbf{70}, 093013 (2004).
\bibitem {Berger1}
E. L.~Berger, Phys. Rev. D\textbf{26}, 105 (1982).
\bibitem{nam}
S.-i Nam, Phys. Rev. D\textbf{86}, 074005 (2012).
\bibitem{watt}
A.D. Martin, W. J. Stirling, R.S. Thorne, G. Watt, Eur. Phys. J.
C\textbf{63}, 189 (2009).

\end{thebibliography}
\end{document}